\documentclass[twocolumn,aps]{revtex4}

\usepackage{graphicx}

\newcommand{\bphi}{{\mbox{\boldmath $\varphi$}}}

\begin{document}

\title{Circuit theory for decoherence in superconducting charge qubits}

\author{Guido Burkard}
\affiliation{IBM T.\ J.\ Watson Research Center,
        P.\ O.\ Box 218,
        Yorktown Heights, NY 10598, USA}

\begin{abstract}
Based on a network graph analysis of the underlying circuit,
a quantum theory of arbitrary superconducting charge qubits is derived.
Describing the dissipative elements of the circuit with a Caldeira-Leggett model,
we calculate the decoherence and leakage rates of a charge qubit.
The analysis includes decoherence due to a dissipative circuit element 
such as a voltage source or the quasiparticle resistances of the Josephson junctions
in the circuit.  The theory presented here is dual to the quantum circuit theory
for superconducting flux qubits.  In contrast to spin-boson models, 
the full Hilbert space structure of the qubit and its coupling
to the dissipative environment is taken into account.
Moreover, both self and mutual inductances of the circuit are fully included.
\end{abstract}

\maketitle

\section{Introduction}
\label{intro}
\begin{figure}[b]
\centerline{\includegraphics[width=6cm]{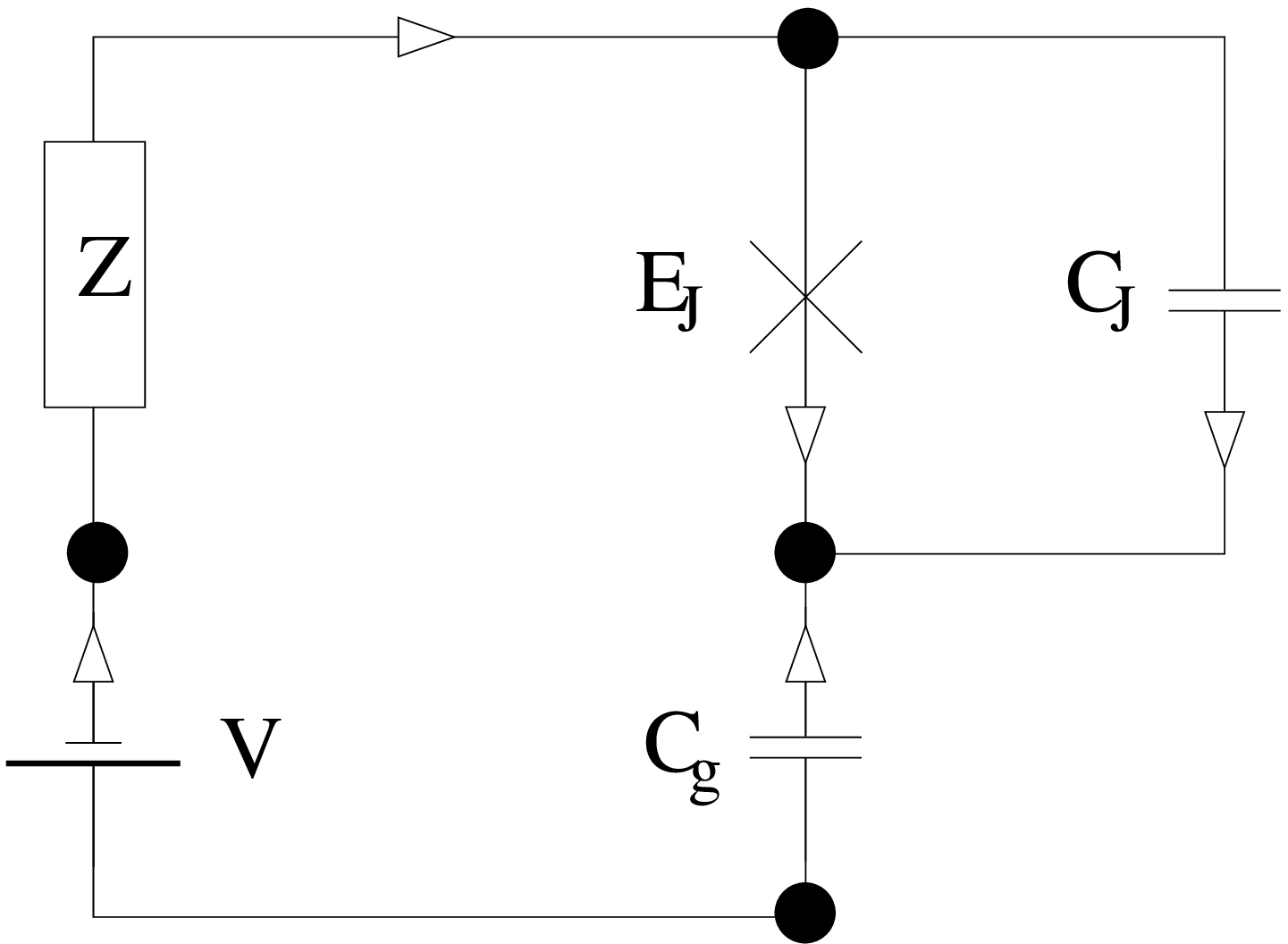}}
  \caption{Circuit graph of a single voltage-biased charge box.
Branches represent a Josephson junction ($E_J$), capacitances
($C_J$ and $C_g$), a voltage source $V$, and the impedance $Z$.
The nodes are shown as black dots;  the node
connecting the junction ($E_J$) to the gate capacitance $C_g$
represents the SC island.}
\label{fig:chbox}
\end{figure}

Various types of quantum bits with Josephson junctions 
in superconducting (SC) circuits are now investigated in theoretical and 
experimental studies \cite{Makhlin01,Devoret04}.
The two types of macroscopic SC qubits, 
the charge \cite{Shnirman97,Averin98,Makhlin99,Nakamura99,Vion02,Pashkin03}
and flux \cite{Mooij99,Orlando,vanderWal,CNHM} qubits,
are distinguished by the relative size of the charging energy $E_C$ and the
Josephson energy $E_J$ of their junctions \cite{fn-phase}.
In flux qubits, also known as
persistent-current qubits, the Josephson energy dominates, $E_J\gg E_C$,
and the state of the qubit is represented as the orientation of a persistent
current in a SC loop \cite{Mooij99,Orlando,vanderWal,CNHM}.  
In contrast to flux qubits, charge 
qubits operate in the regime $E_C \gg E_J$, and are represented 
as the charge state of a small SC island
(presence, $|1\rangle$, or absence, $|0\rangle$, of an 
extra Cooper pair) which is capacitively coupled to 
SC leads \cite{Shnirman97,Averin98,Makhlin99,Nakamura99,Vion02,Pashkin03}
(Fig.~\ref{fig:chbox}).  The quantronium \cite{Vion02} is a charge qubit
that operates in a regime close to $E_C \approx E_J$.

Both types of SC qubits suffer from decoherence that is caused
by a several sources.  In flux qubits, the Johnson-Nyquist noise from lossy
circuit elements (e.g., current sources) has been identified as one important
cause of decoherence \cite{Tian99,Tian02,WWHM}.  
A systematic theory of decoherence of a qubit from such
dissipative elements, based on the network graph analysis \cite{Devoret} 
of the underlying SC circuit, was developed for SC flux qubits \cite{BKD},
and successfully applied to study the effect of asymmetries in a persistent-current
qubit \cite{asym}.  Decoherence in charge qubits has previously been investigated
using the spin-boson model \cite{Makhlin01,Makhlin04}.

Here, we develop a general network graph theory for charge qubits and give examples
for its application.  As in the case of the circuit theory for flux qubits,
we are not restricted to a Hilbert space of the SC device which is 
\textit{a priori} truncated to two levels only.  In other words, in contrast 
to the spin-boson model,
our theory is capable of describing \textit{leakage} errors \cite{Fazio99},
i.e., unwanted
transitions to states that are outside the subspace spanned by the logical
qubit states $|0\rangle$ and $|1\rangle$.  The description presented here
is an extension of earlier results on the SC flux qubits \cite{BKD} and
has potential applications to hybrid charge-flux qubits \cite{Vion02}.
The role of the self and mutual inductances in SC charge qubits have been 
previously studied \cite{You01}, in particular as a means of coupling
two SC charge qubits \cite{Makhlin01,Makhlin04}.  Here, we fully 
and systematically take into account
self and mutual inductances in the underlying SC circuit.

While the circuit theory developed in Secs.~\ref{graphtheory}, \ref{eqmot}, 
and \ref{quantumtheory} can be applied to any SC charge qubit,  
its usefulness will be illustrated with some specific examples 
of charge qubit circuits that have been studied before in Sec.~\ref{examples},
where we reproduce and extend some previously known results.
However, we stress that the circuit theory results are more general than
previously applied methods for the following reasons.
(i) The derived Hamiltonian is not 
\textit{a priori} truncated to a two-dimensional subspace, which allows us to
treat leakage and to \textit{derive} the matrix element of the system-bath coupling.
(ii) The capacitance matrix of the circuit is fully taken into account, and
no assumption about the relative magnitude of gate and Josephson capacitances
has to be made.
(iii) The inductance matrix of the circuit is fully taken into account.

Any number of dissipative elements $Z$ (external impedances, resistances) can
be included in the circuit theory.  In our treatment of the system-bath 
Hamiltonian and the decoherence and relaxation rates in Sec.~\ref{quantumtheory},
we choose to restrict ourselves to the case of a single impedance $Z$ in
order to keep the notation simple.  However, the analysis can readily be
extended to multiple impedances in analogy to SC flux qubits \cite{BB}.

\section{Network graph theory}
\label{graphtheory}
The purpose of this Section is to derive Kirchhoff's laws and the
current-voltage relations (CVRs) for the circuit of a general SC charge
qubit in an appropriate form for their later use in the derivation
of the classical equations of motion of the circuit (Sec.~\ref{eqmot}).

Our analysis (see also \cite{BKD}) starts with the representation of the 
SC circuit as a directed graph, in which the
branches $b_1,b_2,\ldots,b_B$ represent one of the following lumped circuit 
elements: a Josephson junction, inductance, capacitance, voltage source, or an external
impedance (e.g., a resistance).
The circuit graph of a single, voltage-biased charge-box in
Fig.~\ref{fig:chbox} is a simple example of a circuit graph.
In our examples, we neglect the quasiparticle resistance of the junctions 
because it makes the analysis simpler and because 
they are typically less important than the impedances of the 
voltage sources;  however, the shunt resistances can easily be included
as additional impedances in the circuit.
The next step is to find a \textit{tree} of the graph, i.e., a loop-free
subgraph which connects all nodes (for each connected piece of the graph,
we choose a connected sub-tree).
The branches $f_1,f_2,\ldots,f_F$ outside the tree are the so-called \textit{chords};
each chord $f_i$, when added to the tree, gives rise to a unique loop,
a \textit{fundamental loop} ${\cal F}_i$ of the circuit.  The topological information
about the graph which is of importance for our analysis can be represented
in the fundamental loop matrix ($i=1,\ldots F$; $j=1,\ldots, B$), 
\begin{equation}
  {\bf F}^{(L)}_{ij} = \left\{\begin{array}{l l}
      1, & \mbox{if $b_j \in {\cal F}_i$ (same direction)},\\
     -1, & \mbox{if $b_j \in {\cal F}_i$ (opposite direction)},\\
      0, & \mbox{if $b_j \notin {\cal F}_i$},
\end{array}\right. 
\end{equation}
where the direction of the fundamental loop ${\cal F}_i$ is
defined to be opposite to the direction of its defining chord $f_i$.
Accordingly, the currents ${\bf I}=(I_1,\ldots,I_B)$
and voltages ${\bf V}=(V_1,\ldots,V_B)$ associated with the 
branches $1,\ldots,B$ of the graph are split up into tree and chord currents
and voltages,
\begin{equation}
  \label{tree-chord-separation}
  {\bf I} = ({\bf I}_{\rm tr}, {\bf I}_{\rm ch}),\quad\quad\quad
  {\bf V} = ({\bf V}_{\rm tr}, {\bf V}_{\rm ch}).
\end{equation}
With this ordering, the fundamental loop matrix assumes the form
\begin{equation}
  \label{BF}
  {\bf F}^{(L)} = \left(- {\bf F}^T \, |\, \openone \right),
\end{equation}
and we will simply refer to the matrix ${\bf F}$ in the following.
Using Eq.~(\ref{BF}), we write Kirchhoff's laws in the following
useful form  \cite{BKD},
\begin{eqnarray}
  {\bf F}  {\bf I}_{\rm ch} &=& - {\bf I}_{\rm tr},\label{IVct-1}\\
  {\bf F}^T{\bf V}_{\rm tr} &=&   {\bf V}_{\rm ch} - {\bf \dot\Phi}_x,\label{IVct-2}
\end{eqnarray}
where ${\bf \Phi}_x=(\Phi_1,\ldots,\Phi_F)$ denote the externally applied magnetic
fluxes threading loops $1,\ldots,F$ of the circuit.
The partition of branch types into tree and chord branches is \textit{dual} to
the flux qubit case \cite{BKD}, i.e., the roles of tree and chord branches are
interchanged.  

Before we proceed, we summarize the assumptions about the circuit
that will be used in the following.
\renewcommand{\labelenumi}{(\roman{enumi})}
\begin{enumerate}
\item \label{asm1} There are no loops containing nothing else than Josephson
junctions (J), external impedances (Z), and voltage sources (V).
This assumption is physically motivated because all loops have
a finite self-inductance.
\item \label{asm2} Voltage sources (V) and impedances (Z) are not inductively
shunted.
\item \label{asm3} There are sufficiently many capacitors (C) in the circuit to independently
shunt all inductors.  A more precise form of this requirement is that
the capacitance matrix ${\cal C}$ has full rank (see below).
\end{enumerate}

Using assumption (i), we may split up the current and voltage 
vectors as
\begin{eqnarray}
  \label{IVsplit}
  {\bf I}_{\rm tr}  =  ({\bf I}_J, {\bf I}_L, {\bf I}_V, {\bf I}_Z), & & 
  {\bf I}_{\rm ch}  =  ({\bf I}_{C_J}, {\bf I}_C, {\bf I}_K), \\
  {\bf V}_{\rm tr}  =  ({\bf V}_J, {\bf V}_L, {\bf V}_V, {\bf V}_Z), & &
  {\bf V}_{\rm ch}  =  ({\bf V}_{C_J}, {\bf V}_C, {\bf V}_K).
\quad\quad \label{IVsplit2}
\end{eqnarray}
The chord current and voltage vectors ${\bf I}_{\rm ch}$
and ${\bf V}_{\rm ch}$ in Eqs.~(\ref{IVsplit}) and (\ref{IVsplit2}) 
contain the branch currents and voltages
of the capacitors ($C_J$,$C$) and chord inductors ($K$);
the tree current and voltage vectors ${\bf I}_{\rm tr}$ and
${\bf V}_{\rm tr}$ contain the branch currents and voltages
of tree inductors ($L$), Josephson junctions ($J$),
external impedances ($Z$), and bias voltage sources ($V$) \cite{fn1}.
The loop matrix ${\bf F}$ then acquires the block form,
\begin{eqnarray}
  \label{Fsplit}
  {\bf F} = \left(\begin{array}{c c c}
      \openone & {\bf F}_{JC} & {\bf F}_{JK} \\
      {\bf 0}  & {\bf F}_{LC} & {\bf F}_{LK} \\
      {\bf 0}  & {\bf F}_{VC} & {\bf F}_{VK} \\
      {\bf 0}  & {\bf F}_{ZC} & {\bf F}_{ZK} \\
\end{array}\right).
\end{eqnarray}
The form of the first column in Eq.~(\ref{Fsplit}) reflects
the fact that the $C_J$ capacitances are (by definition) shunted in parallel to the
Josephson junctions.  Moreover, assumption (ii) above
implies ${\bf F}_{VK}={\bf F}_{ZK}={\bf 0}$.
In order to derive the equations of motion, we formally define the branch 
charges and fluxes ($X=C, K, J, L, Z, V$),
\begin{eqnarray}
  {\bf I}_X(t) &=& \dot{\bf Q}_X(t),    \label{charges}\\
  {\bf V}_X(t) &=& \dot{\bf \Phi}_X(t). \label{fluxes}
\end{eqnarray}
where the formal fluxes of the Josephson branches are the
SC phase differences across the junctions, according to the 
second Josephson relation,
\begin{equation}
\label{Josephson-2}
  \frac{{\bf \Phi}_J}{\Phi_0} =  \frac{\bphi}{2\pi},
\end{equation}
with $\Phi_0=h/2e$.
The current-voltage relations (CVRs) of the Josephson, capacitance, and 
external impedance branches are
\begin{eqnarray}
  {\bf I}_J  &=&  {\bf I}_{\rm c} \,\mbox{\boldmath $\sin$} \bphi
    =   {\bf I}_{\rm c} \,\mbox{\boldmath $\sin$} \left(2\pi\frac{{\bf \Phi}_J}{\Phi_0}\right) ,\label{CVR-J}\\
  {\bf Q}_C  &=&  {\bf C}{\bf V}_C,\label{CVR-C}\\
  {\bf V}_Z  &=&  {\bf Z} * {\bf I}_Z\label{CVR-Z},
\end{eqnarray}
where the convolution is defined as
$({\bf f} * {\bf g}) (t) = \int_{-\infty}^t {\bf f}(t - \tau) {\bf g}(\tau) d\tau $.
The CVR for the inductive branches has the following matrix form,
\begin{equation}
  \label{inductance-1}
   \left(\begin{array}{c}{\bf \Phi}_L\\ {\bf \Phi}_K\end{array}\right)
  =\left(\begin{array}{l l}{\bf L}        & {\bf L}_{LK}\\
                           {\bf L}_{LK}^T & {\bf L}_{K} \end{array}\right)
   \left(\begin{array}{c}{\bf I}_L\\ {\bf I}_K\end{array}\right)
  \equiv {\bf L}_{\rm t} \left(\begin{array}{c}{\bf I}_L\\ {\bf I}_K\end{array}\right),
\end{equation}
where ${\bf L}$ and ${\bf L}_K$ are the self inductances of the chord and tree branch
inductors, resp., off-diagonal elements describing the mutual inductances among
chord inductors and tree inductors separately, and  ${\bf L}_{LK}$ is the
mutual inductance matrix between tree and chord inductors.
Since the total inductance matrix is symmetric and positive, i.e.\
${\bf v}^T {\bf L}_{\rm t} {\bf v}>0$ for all real vectors ${\bf v}$,
its inverse exists, and we find
\begin{eqnarray}
   \left(\begin{array}{c}{\bf I}_L\\ {\bf I}_K\end{array}\right)
  &=&\left(\begin{array}{c c}\bar{\bf L} ^{-1}    & -{\bf L}^{-1}{\bf L}_{LK}\bar{\bf L}_K^{-1}\\
                   -{\bf L}_K^{-1}{\bf L}_{LK}^T\bar{\bf L}^{-1} & \bar{\bf L}_{K} ^{-1} \end{array}\right)
   \left(\begin{array}{c}{\bf \Phi}_L\\ {\bf \Phi}_K\end{array}\right)\nonumber\\
  &\equiv&{\bf L}_{\rm t}^{-1}\left(\begin{array}{c}{\bf \Phi}_L\\ {\bf \Phi}_K\end{array}\right)
\label{inductance-2}
\end{eqnarray}
with the definitions
\begin{eqnarray}
  \bar{\bf L}        &=& {\bf L}   - {\bf L}_{LK}   {\bf L}_K^{-1} {\bf L}_{LK}^T,\label{Lbar}\\
  \bar{\bf L}_{K}    &=& {\bf L}_K - {\bf L}_{LK}^T {\bf L}  ^{-1} {\bf L}_{LK}.\label{LKbar}
\end{eqnarray}

\section{Classical equation of motion}
\label{eqmot}

In this Section, we derive the classical equation of
motion of the dynamical variables ${\bf \Phi} = ({\bf \Phi}_J , {\bf \Phi}_L)$
of the circuit.

We now combine Kirchhoff's laws, Eqs.~(\ref{IVct-1}) and (\ref{IVct-2}), 
and the the CVRs, Eqs.~(\ref{CVR-J})--(\ref{LKbar}), in order
to derive the classical equations of motion of the circuit.  These will
then be used in Sec.~\ref{quantumtheory} to find the Hamiltonian
suitable for quantization.  The details of the derivation are explained
in Appendix \ref{deriv-eqmot}.

Equations (\ref{eqm-3}) and (\ref{eqm-6}) can be summarized as
\begin{equation}
  \label{summary1}
  {\cal C}\dot{\bf \Phi} = {\bf Q} -{\bf C}_V{\bf V}-{\cal F}_{C}{\bf C}_Z * {\bf V}_C,
\end{equation}
with the combined flux vector 
${\bf \Phi} = ({\bf \Phi}_J , {\bf \Phi}_L)= (\Phi_0 \bphi / 2\pi , {\bf \Phi}_L)$,
and the canonical charge
\begin{equation}
  \label{Q}
  {\bf Q} = - \left(\begin{array}{c}{\bf Q}_{J}\\{\bf Q}_{L}\end{array}\right)
            - {\cal F}_{K} {\bf Q}_K.
\end{equation}
Note that in the SC charge qubits studied in 
Ref.~\onlinecite{Nakamura99}, the Josephson junctions lead to (otherwise
only capacitively coupled) SC islands, with the consequence that there are no 
chord inductors (K), and ${\bf Q} = - ({\bf Q}_{J}, {\bf Q}_{L})^T$.
However, the quantronium circuits \cite{Vion02} 
which have hybrid charge and flux nature, cannot be described without
chord inductors.  In the following, we will derive our theory for the 
most general case including chord inductors, but further below, we will 
also discuss the much simpler special case without chord inductors.
In Eqs.~(\ref{summary1}) and (\ref{Q}) we have also introduced the notation
\begin{equation}
{\cal F}_{X}=\left(\begin{array}{c}{\bf F}_{JX}\\{\bf F}_{LX}\end{array}\right),
\end{equation}
for $X=C,K$, and 
the capacitance matrices
\begin{eqnarray}
  \label{eq:C}
  {\cal C} &=& \left(\begin{array}{c c}
  {\bf C}_{\rm tot}    & {\bf C}_{JL}\\
  {\bf C}_{JL}^T       & {\bf C}_L
  \end{array}\right)
\equiv\left(\begin{array}{c c}
  {\bf C}_J & 0 \\
  0         & 0
  \end{array}\right) 
  + {\cal F}_{C}{\bf C}{\cal F}_{C}^T,\quad\quad\\
  {\bf C}_V &=& \left(\begin{array}{c}
  {\bf C}_{JV}\\
  {\bf C}_{LV}
  \end{array}\right)
\equiv {\cal F}_{C} {\bf C} {\bf F}_{VC}^T, \label{eq:CV}\\
  \label{eq:CZ}
  {\bf C}_Z(\omega) &=& i\omega {\bf C}{\bf F}_{ZC}^T {\bf Z}(\omega) {\bf F}_{ZC}{\bf C}.
\end{eqnarray}

We can further rewrite the dissipative term in Eq.~(\ref{summary1}) by using
Eq.~(\ref{IVct-2}) (capacitance part),
solving for ${\bf V}_C$, and substituting the solution back
into Eq.~(\ref{summary1}), with the result
\begin{equation}
 \left({\cal C} + {\cal C}_Z\right)*\dot{\bf \Phi}  = {\bf Q} - {\bf C}_V{\bf V}, \label{eq:firstorder-1}
\end{equation}
where we have introduced
\begin{eqnarray}
{\cal C}_Z(\omega) &=& \bar{\bf m}\bar{\bf C}_Z(\omega) \bar{\bf m}^T,\label{eq:diss-2}\\
\bar{\bf m} &=& {\cal F}_{C}{\bf C}{\bf F}_{ZC}^T
            = \left(\begin{array}{c} \bar{\bf m}_{J} \\ \bar{\bf m}_{L}\end{array}\right)\label{mbar}\\
  \bar{\bf C}_Z(\omega)   &=& i\omega {\bf Z}(\omega)\left(\openone + {\bf F}_{ZC}{\bf C}{\bf F}_{ZC}^T i\omega{\bf Z}(\omega)\right)^{-1}.\label{CZ}
\end{eqnarray}
Using the symmetry of ${\bf C}_Z(\omega)$, we can show that $\bar{\bf C}_Z(\omega)$ is also a symmetric matrix.

We obtain the equation of motion from Eq.~(\ref{eq:firstorder-1}) by taking
the derivative with respect to time, and using Eq.~(\ref{charges}) with $X=K,L$
and Eq.~(\ref{inductance-2}),
\begin{equation}
  \label{eq:Hamilton-Q}
  \left({\cal C} + {\cal C}_Z\right)*\ddot{\bf \Phi} = \dot{\bf Q}  = -\frac{\partial U}{\partial {\bf \Phi}},
\end{equation}
with the potential 
\begin{equation}
  \label{eq:U}
  U({\bf \Phi})   = -{\bf L}_J^{-1} \,\mbox{\boldmath $\cos$} \bphi
                  +\frac{1}{2}{\bf \Phi}^T {\bf M}_0 {\bf \Phi}
                  +{\bf \Phi}^T {\bf N}{\bf \Phi}_x,
\end{equation}
where ${\bf \Phi} = (\Phi_0\bphi/2\pi,{\bf \Phi}_L)$ and
\begin{equation}
  \label{MN}
  {\bf M}_0 = {\cal G}{\bf L}_t^{-1} {\cal G}^T,\quad\quad\quad
  {\bf N}   = {\cal G}{\bf L}_t^{-1} \left(\begin{array}{c c} 0 & \openone_K \end{array}\right)^T,
\end{equation}
with the $(N_L+N_K)\times (N_J+N_L)$ block matrix
\begin{equation}
  \label{G}
  {\cal G} = \left(\begin{array}{c c}
0          & -{\bf F}_{JK}\\
\openone_L & -{\bf F}_{LK}
\end{array}\right).
\end{equation}
Using ${\bf L}_t^T={\bf L}_t$, we observe that ${\bf M}_0^T={\bf M}_0$.
In the absence of chord inductors (K), we find
${\bf \Phi}^T {\bf M}_0 {\bf \Phi} = {\bf \Phi}_L^T {\bf L}^{-1} {\bf \Phi}_L$
and ${\bf N}=0$, whereas
in the absence of tree inductors (L), we obtain
$\frac{1}{2}{\bf \Phi}^T {\bf M}_0 {\bf \Phi} + {\bf \Phi}^T {\bf N}{\bf \Phi}_x = \frac{1}{2}\left({\bf F}_{JK}^T \bphi + {\bf \Phi}_x \right)^T  {\bf L}_K^{-1} \left({\bf F}_{JK}^T \bphi + {\bf \Phi}_x \right) +{\rm const}$.

By bringing the dissipative term in Eq.~(\ref{eq:Hamilton-Q})
to the right hand side and using assumption (iii), 
we find the equation of motion
\begin{equation}
  \label{eq:eq-mot}
  {\cal C}\ddot{\bf \Phi}  = -\frac{\partial U}{\partial {\bf \Phi}} 
                             - {\cal C}_d*{\cal C}^{-1}\dot{\bf Q},
\end{equation}
with the dissipation matrix
\begin{equation}
  \label{eq:Cd}
  {\cal C}_d(\omega) =      \left(1+{\cal C}_Z(\omega){\cal C}^{-1}\right)^{-1} {\cal C}_Z(\omega)
                     \equiv \bar{\bf m} {\bf K}(\omega) \bar{\bf m}^T,
\end{equation}
and the frequency-dependent kernel
\begin{equation}
  \label{eq:K}
  {\bf K}(\omega) = \bar{\bf C}_Z(\omega)\left( \openone 
                       + \bar{\bf m}^T{\cal C}^{-1}\bar{\bf m}\bar{\bf C}_Z(\omega)\right)^{-1}.
\end{equation}
Since both $\bar{\bf C}_Z(\omega)$ and ${\cal C}$ are symmetric matrices, we find that
${\bf K}(\omega)$, and thus also ${\cal C}_d(\omega)$, are symmetric.
Moreover, we know that ${\cal C}_d(t)$ inherits two additional properties from ${\bf Z}(t)$:
it is also real and causal, i.e., ${\cal C}_d(t)=0$ for $t<0$.
In a perturbation expansion in ${\bf Z}^2$, the lowest order term in ${\bf K}(\omega)$ 
is simply ${\bf K}(\omega) = i\omega {\bf Z}(\omega) + O({\bf Z})^2$.

In deriving Eq.~(\ref{eq:eq-mot}), we have used assumption (iii) 
that the matrix ${\cal C}$ has full rank, 
such that ${\cal C}^{-1}$ exists.  Since all junctions are capacitively shunted, we know
that ${\bf C}_{\rm tot}$ has full rank, hence $N_J \le {\rm rank}{\cal C} \le N_J+N_L$,
where $N_X$ is the number of branches of type $X$.  The case ${\rm rank}{\cal C} < N_J+N_L$
occurs if there are not sufficiently many capacitances in the circuit to independently
shunt all inductors.  In that case, Eq.~(\ref{eq:eq-mot}), without the dissipative part, 
contains $l=N_J+N_L-{\rm rank}{\cal C}$ constraints that can be used to eliminate $l$
degrees of freedom.  In the case of SC flux qubits \cite{BKD}, it was assumed
that only the junctions are shunted by capacitors (${\rm rank}{\cal C}=N_J$), 
thus $l$ is the number of tree inductors.

\section{Quantum theory}
\label{quantumtheory}

The purpose of this section is to derive the Hamiltonian
of the circuit, including its dissipative elements, and then 
to quantize this Hamiltonian in order to have a description 
of the quantum dissipative dynamics of the circuit from which
a master equation and, finally, the decoherence rates can be
derived.

The Hamiltonian of the circuit
\begin{equation}
  \label{eq:HS}
  {\cal H}_S = \frac{1}{2}\left({\bf Q} - {\bf C}_V{\bf V}\right)^T{\cal C}^{-1}\left({\bf Q} 
                                  - {\bf C}_V{\bf V}\right) + U({\bf \Phi}),
\end{equation}
giving rise to the equation of motion (\ref{eq:eq-mot}) without 
dissipation (${\bf Z}=0$),
can readily be quantized with the commutator rule
\begin{equation}
  \label{commutator}
  \left[\Phi_i,Q_j\right]=i\hbar\delta_{ij}.
\end{equation}
A somewhat subtle point here is that while the inductor flux variables ${\bf \Phi}_L$
are defined on an infinite domain, the Josephson flux variables
${\bf \Phi}_J = (\Phi_0/2\pi)\bphi$ are defined on a compact domain
since they are periodic with period $\Phi_0$.  Upon imposing
Eq.~(\ref{commutator}), this leads to charge operators ${\bf Q}_L$ with a
continuous spectrum and ${\bf Q}_J$ with
a discrete spectrum with eigenvalues $Q_{Ji} = 2en_i$, with $n_i$ integer~\cite{Devoret04}.

In order to describe the dissipative dynamics of the SC circuit,
we construct a Caldeira-Leggett Hamiltonian \cite{CaldeiraLeggett}
${\cal H} = {\cal H}_S + {\cal H}_B + {\cal H}_{SB}$
that reproduces the classical dissipative equation of motion Eq.~(\ref{eq:eq-mot}).
For simplicity, we will restrict ourselves to the case of
a single impedance $Z$ here, where a single bath of harmonic oscillators
can be used to model the dissipative environment,
\begin{equation}
  \label{eq:HB}
  {\cal H}_{B} = \sum_\alpha \left(\frac{p_{\alpha}^2}{2 m_{\alpha}}
                                   +\frac{1}{2}m_{\alpha} \omega_{\alpha}^2 x_{\alpha}^2\right).
\end{equation}
We choose the system-bath coupling to be of the form
\begin{equation}
  \label{eq:HSB}
  {\cal H}_{SB} = {\cal C}^{-1}\bar{\bf m}\cdot{\bf Q} \sum_\alpha c_\alpha x_\alpha
                = \bar{\bf m}\cdot  {\cal C}^{-1}{\bf Q} \sum_\alpha c_\alpha x_\alpha,
\end{equation}
such that it reproduces the classical equation of motion Eq.~(\ref{eq:eq-mot}),
with a spectral density of the bath modes (for a derivation, see Appendix \ref{CLderivation})
\begin{equation}
  J(\omega) =  - {\rm Im}\, K(\omega) \label{eq:JK}.
\end{equation}
Note that the kernel $K$ has become a scalar because we are now only
dealing with a single external impedance.

From the Hamiltonian $\cal H$, the master equation for the evolution of the system
density matrix can be derived \cite{BKD}.  In the Born-Markov approximation,
the matrix elements ${\rho}_{nm}=\langle n|\rho_S|m\rangle$,
where ${\cal H}_S|n\rangle=\omega_n|n\rangle$,
obey the Redfield equation \cite{Redfield}
\begin{equation}
  \dot{\rho}_{nm}(t) 
   = -i\omega_{nm}\rho_{nm}(t) -\sum_{kl}R_{nmkl}\rho_{kl}(t),\label{Redfield-equation}
\end{equation}
with $\omega_{nm}=\omega_n-\omega_m$, and with the Redfield tensor,
\begin{equation}
  R_{nmkl} = \delta_{lm}\!\sum_r \Gamma_{nrrk}^{(+)} + \delta_{nk}\!\sum_r \Gamma_{lrrm}^{(-)}
-\Gamma_{lmnk}^{(+)}-\Gamma_{lmnk}^{(-)},\label{RGamma}
\end{equation}
where $(\Gamma_{lmnk}^{(+)})^* = \Gamma_{knml}^{(-)}$, and
\begin{eqnarray}
{\rm Re}\Gamma_{lmnk}^{(+)} &=&  \frac{1}{\hbar}({\bf m}\cdot{\bf Q})_{lm} ({\bf m}\cdot{\bf Q})_{nk} 
J(|\omega_{nk}|)\frac{e^{-\hbar \beta \omega_{nk}/2}}{\sinh \hbar \beta|\omega_{nk}|/2}\, ,\nonumber\\
{\rm Im}\Gamma_{lmnk}^{(+)} &=& -\frac{1}{\hbar}({\bf m}\cdot{\bf Q})_{lm} ({\bf m}\cdot{\bf Q})_{nk} \times \label{Gp} \\
&& \times \frac{2}{\pi} P\!\!\int_0^\infty \!\!\!\!\!\!d\omega \frac{J(\omega)}{\omega^2 \!-\!\omega_{nk}^2}\!\left(\!\omega\!-\!\omega_{nk}\coth \frac{\hbar \beta\omega}{2}\!\right), \nonumber
\end{eqnarray}
and ${\bf m}={\cal C}^{-1}\bar{\bf m}$.

The Redfield equation (\ref{Redfield-equation}) can be derived for arbitrary
SC circuits. The SC circuit can represent a single qubit or a number of qubits.
In order to make connection with single-qubit experiments, we apply the theory to
the case of a SC circuit representing a single qubit.
Restricting ourselves to the two lowest levels and working in the
secular approximation \cite{BKD}, the Redfield equation Eq.~(\ref{Redfield-equation})
turns into a Bloch equation with the relaxation ($T_1$) and decoherence ($T_2$) times,
\begin{eqnarray}
  \frac{1}{T_1} &=& \frac{4}{\hbar}|\langle 0|{\bf m}\cdot{\bf Q}|1\rangle|^2 J(\omega_{01}) \coth\frac{\hbar\omega_{01}}{2k_B T}, \label{T1}\\
  \frac{1}{T_2} &=& \frac{1}{2 T_1} + \frac{1}{T_\phi},\label{T2}\\\
  \frac{1}{T_\phi} &=&  \frac{1}{\hbar}|\langle 0|{\bf m}\cdot{\bf Q}|0\rangle-\langle 1|{\bf m}\cdot{\bf Q}|1\rangle|^2 \left.\frac{J(\omega)}{\hbar\omega}\right|_{\omega\rightarrow 0} \!\!\!\!\!\!\!\!\! 2k_B T. \quad\quad\label{Tphi}
\end{eqnarray}
In the semiclassical approximation \cite{BKD}, 
$\langle 0|{\bf Q}|1\rangle\approx (1/2)(\Delta/\omega_{01})\Delta{\bf Q}$ and
$\langle 0|{\bf Q}|0\rangle - \langle 1|{\bf Q}|1\rangle\approx (\epsilon/\omega_{01})\Delta{\bf Q}$,
where $\Delta{\bf Q}={\bf Q}_0-{\bf Q}_1$ is the ``distance'' between  two localized low-energy 
classical charge 
states ${\bf Q}_0$ and ${\bf Q}_1$, $\epsilon$ is the classical energy difference 
and $\Delta$ the tunneling amplitude between them,
and $\omega_{01}=\sqrt{\Delta^2+\epsilon^2}$ is the energy splitting between the two 
quantum eigenstates in this energy double well.
Within this approximation, we find
\begin{eqnarray}
  \frac{1}{T_1} &=& \frac{1}{\hbar}|{\bf m}\cdot \Delta{\bf Q}|^2 \left(\frac{\Delta}{\omega_{01}}\right)^2 J(\omega_{01}) \coth\frac{\hbar\omega_{01}}{2k_B T}, \label{T1-sc}\\
  \frac{1}{T_\phi} &=&  \frac{1}{\hbar}|{\bf m}\cdot \Delta{\bf Q}|^2 \left(\frac{\epsilon}{\omega_{01}}\right)^2 \left.\frac{J(\omega)}{\hbar\omega}\right|_{\omega\rightarrow 0} \!\!\!\!\!\!\!\!\! 2k_B T. \quad\quad\label{Tphi-sc}
\end{eqnarray}
The leakage rates from the logical state $k=0,1$ to states $n=2,3,\ldots$ outside
the computational subspace can be estimated as
\begin{equation}
  \label{leakage}
  \frac{1}{T_L} = \frac{4}{\hbar}\sum_{n=2}^\infty|\langle k|{\bf m}\cdot{\bf Q}|n\rangle|^2 J(\omega_{nk}) \coth\frac{\hbar\omega_{nk}}{2k_B T}.
\end{equation}

\section{Examples}
\label{examples}

\subsection{Single Charge Box}
\label{ssec-chbox}

The voltage-biased charge box is shown in Fig.~\ref{fig:chbox},
where the inductance of the leads has been neglected for simplicity (no $L$ and $K$ branches).
The tree of the graph is given by the Josephson, voltage source,
and impedance branches.
For the loop matrices, we simply find
\begin{equation}
  \label{eq:chbox-FJC}
  {\bf F}_{JC} = {\bf F}_{VC} = {\bf F}_{ZC} = 1.
\end{equation}
With the capacitances
\begin{equation}
  {\cal C} \equiv C_{\rm tot} = C_J+C_g,\quad\quad
  C_V = C_g,
\end{equation}
we arrive at the Hamiltonian,
\begin{equation}
  \label{eq:chbox-HS}
  {\cal H}_S = \frac{(Q_J+C_g V)^2}{2C_{\rm tot}}+E_J\cos\varphi .
\end{equation}
The coupling to the environment is characterized by
${\bf m}=( C_g/C_{\rm tot})$.
As an example, we give here the relaxation and dephasing times,
with $m=|{\bf m}|=C_g/C_{\rm tot}$,
\begin{eqnarray}
  \frac{1}{T_1} &=& 2\pi m^2 4 |\langle 0|n|1\rangle|^2 \frac{4{\rm Re}Z(\omega_{01})}{R_Q} \omega_{01} \coth\frac{\hbar\omega_{01}}{2k_B T}, \label{T1-chbox}\\
  \frac{1}{T_\phi} &=& 2\pi m^2 |\langle 0|n|0\rangle-\langle 1|n|1\rangle|^2 \frac{4{\rm Re}Z(0)}{R_Q}  \frac{2k_B T}{\hbar}, \quad\quad\label{Tphi-chbox}
\end{eqnarray}
where $n=Q/2e$ and $R_Q=h/e^2$.
In the semiclassical limit, $\langle 0|n|1\rangle\approx (1/2)(\Delta/\omega_{01})\Delta n$
and $\langle 0|n|0\rangle-\langle 1|n|1\rangle\approx (\epsilon/\omega_{01})\Delta n$.
With $\Delta n\approx 1$, we reproduce the results in \cite{Makhlin01}.
Typical leakage rates are of the form of $1/T_1$, with the matrix element replaced by
$|\langle 0|n|k\rangle|$ and $|\langle 1|n|k\rangle|$, where $k\ge 2$ labels a state
other than the two qubit states, and with $\omega_{01}$ replaced by $\omega_{lk}$ ($l=0,1$).

\subsection{Flux-controlled Josephson junction}
\label{ssec-fc}

\begin{figure}
\centerline{\includegraphics[width=7cm]{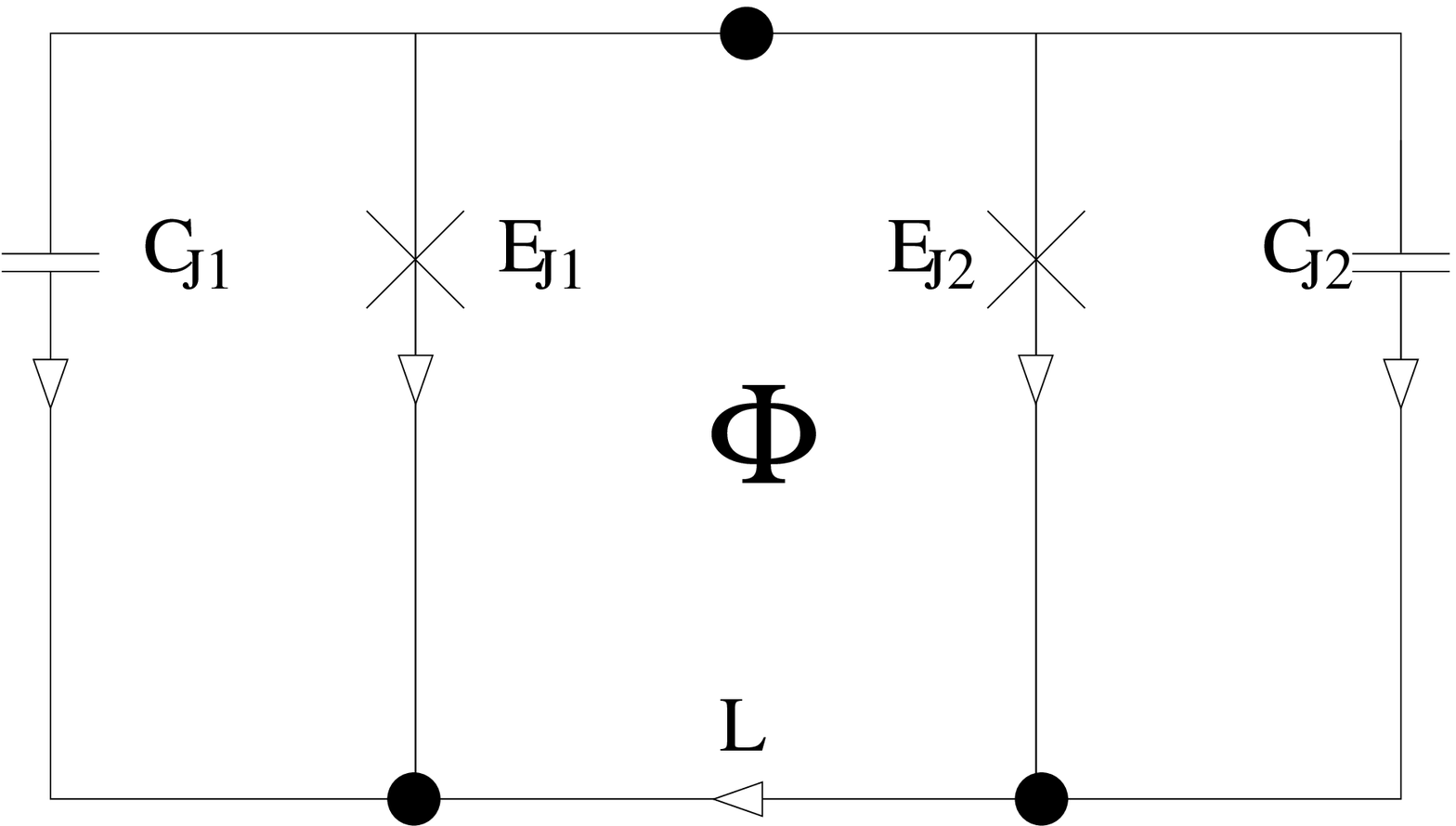}}
  \caption{A flux-controlled Josephson junction.}
  \label{fig:flux-controlled}
\end{figure}
\begin{figure}
\centerline{\includegraphics[width=7cm]{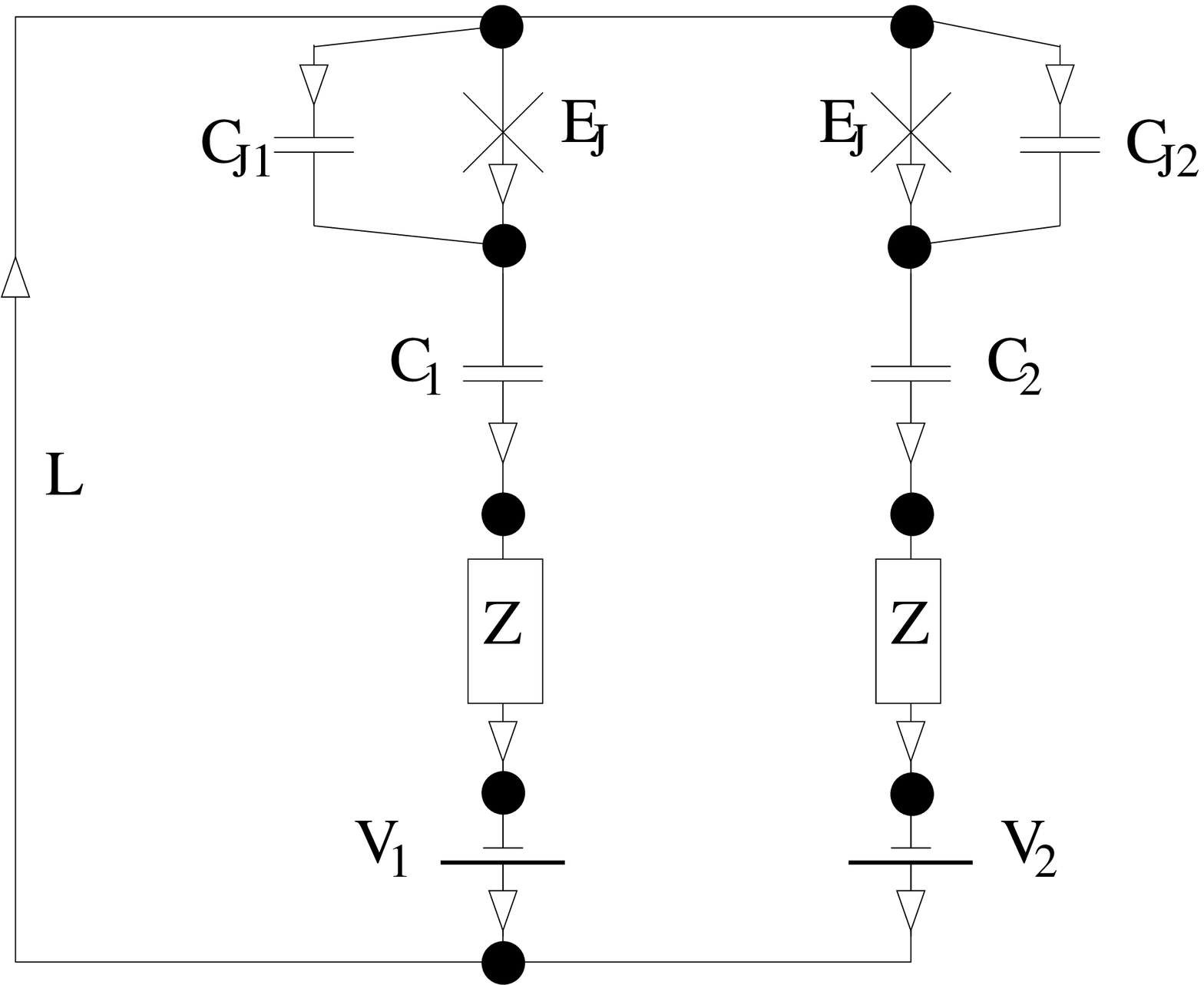}}
  \caption{Two inductively coupled charge boxes.}
  \label{fig:ind-coupl}
\end{figure}
A flux-controlled Josephson junction is a SC loop with two 
junctions  which acts as
an effective Josephson junction with a flux-dependent Josephson 
energy \cite{Makhlin99}.  The circuit Fig.~\ref{fig:flux-controlled}
we use to describe the the flux-controlled junction comprises a
chord inductance ($K$) with inductance $L$.  
The tree consists of the two Josephson branches.
The only relevant loop matrix is
${\bf F}_{JK}=\left(\begin{array}{c c} 1 & -1 \end{array}\right)^T$.
In the limit $L\rightarrow 0$, and if $E_{J1}=E_{J2}$, we find 
${\bf F}_{JK}^T\bphi +\Phi_x = \varphi_1-\varphi_2 + \Phi_x \rightarrow 0$,
which leads us to the Hamiltonian
\begin{equation}
  \label{eq:fc-HS}
  {\cal H}_S = \frac{Q^2}{2 \bar{C}} - E_J(\Phi_x) \cos\varphi,
\end{equation}
where $\varphi = \varphi_1 + \pi \Phi_x/\Phi_0$, $\bar C = C_{J1} + C_{J2}$, and
$E_J(\Phi_x)  =  2 E_J \cos(2\pi\Phi_x/\Phi_0)$.

\subsection{Inductively coupled charge boxes}
\label{ssec-icoupl}
We now turn to the case of two charge boxes of the type discussed in
Sec.~\ref{ssec-chbox}, coupled via an inductive loop \cite{Makhlin01,Makhlin99},
as shown in Fig.~\ref{fig:ind-coupl}.
Here, the tree consists of all Josephson, voltage source, and impedance
branches, plus the inductive branch $L$,
and the loop matrices are
\begin{equation}
\label{eq:icoupl-FJ}
  {\bf F}_{JC} = {\bf F}_{VC} = {\bf F}_{ZC} = 
\left(\begin{array}{c c}
      1 & 0\\
      0 & 1
\end{array}\right),
\:
  {\bf F}_{LC} = \left(\begin{array}{c c}
      1 & 1
\end{array}\right).
\end{equation}
With the two capacitance matrices
${\bf C} = {\rm diag}(C_1,C_2)$ and ${\bf C}_J = {\rm diag}(C_{J1},C_{J2})$,
we find
${\bf C}_{\rm tot} = {\bf C} + {\bf C}_J$,
${\bf C}_{JV} = {\bf C}$, 
${\bf C}_{JL} = {\bf C}_{LV}^T = (C_{1}, C_{2})^T$, and
${\bf C}_L = C_1+C_2$.
The vector $\bar{\bf m}$ consists of the two parts
${\bf m}_J = C$ and
${\bf m}_L = \left(\begin{array}{c c} C_1 & C_2\end{array}\right)$.
With Eq.~(\ref{eq:HS}) and the inverse of the total capacitance matrix,
\begin{widetext}
\begin{equation}
  \label{eq:3}
  {\cal C}^{-1}
= \frac{1}{\gamma}
\left(\begin{array}{c c c}
(C_1 +C_2) C_{J2}-C_2^2 & C_1 C_2 & -C_1 C_{J2}\\
C_1 C_2 & (C_1+C_2)C_{J1}-C_1^2 & -C_2 C_{J1}\\
-C_1 C_{J2} & -C_2 C_{J1} &  C_{J1} C_{J2}
\end{array}\right)
\equiv\left(\begin{array}{c c c}
C_{\rm eff,1}^{-1}    &   C_{\rm eff,12}^{-1}   &   C_{{\rm eff},L1}^{-1} \\
C_{\rm eff,12}^{-1}   &   C_{\rm eff,2}^{-1}    &   C_{{\rm eff},L2}^{-1} \\
C_{{\rm eff},L1}^{-1} &   C_{{\rm eff},L2}^{-1} &   C_{{\rm eff},L}^{-1}
\end{array}\right),
\end{equation}
\end{widetext}
where $\gamma=(C_1+C_2)C_{J1}C_{J2}-C_1^2 C_{J2}-C_2^2 C_{J1}$,
the Hamiltonian of the coupled system can be written as,
\begin{eqnarray}
  {\cal H}_S &=& \sum_{i=1,2}\left(\frac{(Q_{Ji}+C_i V_i)^2}{2 C_{{\rm eff},i}}
                           +E_{Ji}\cos\varphi_i \right) \nonumber\\
                & &        +\frac{(Q_L + C_1 V_1 + C_2 V_2)^2}{2 C_{{\rm eff},L}}
                           +\frac{\Phi_L^2}{2L}    \label{eq:icoupl-HS}\\
                & &        +\frac{(Q_{J1}+C_1 V_1)(Q_{J2}+C_2 V_2)}{C_{\rm eff,12}}\nonumber\\
                & &        -\sum_{i=1,2}\frac{(Q_{Ji}+C_i V_i)(Q_L + C_1 V_1 + C_2 V_2)}{C_{{\rm eff},Li}}.\nonumber
\end{eqnarray}
While the last term in Eq.~(\ref{eq:icoupl-HS}) couples each qubit to the
$LC$ mode associated with the inductor $L$, and is thus responsible for the
inductive coupling of the qubits, the second last term provides a direct
capacitive coupling between the qubits.
In the limit $C_i \ll C_{Ji}$, we reproduce the results of \cite{Makhlin01};
however, there are additional terms of order $C_i/C_{Ji}$, in particular
the new term $\propto 1/C_{\rm eff,12}$ in the Hamiltonian that capacitively couples
the qubits directly.
Since the coupled system involves at least four levels (more if excited states of the $LC$
coupling circuit or higher qubit levels are included), it can no longer be described
by a two-level Bloch equation with parameters $T_1$ and $T_2$.  We can however fix one of the
qubits to be in a particular state, say $|0\rangle$, and then look at the ``decoherence rates''
of the other qubit.
To lowest order in $C_i/C_{Ji}$, these rates due to the impedance $Z_i$ have the form
($q_i = C_i/(C_1+C_2)$)
\begin{eqnarray}
  \frac{1}{T_1} &=& 2\pi q_i^2 4 |\langle 00|n_L|10\rangle|^2 \frac{4{\rm Re}Z_i(\omega_{01})}{R_Q} \omega_{01} \coth\frac{\hbar\omega_{01}}{2k_B T}, \label{T1-icoupl}\\
  \frac{1}{T_\phi} &=& 2\pi q_i^2 |\langle 00|n_L|00\rangle-\langle 10|n_L|10\rangle|^2 \frac{4{\rm Re}Z_i(0)}{R_Q}  \frac{2k_B T}{\hbar}. \quad\quad\label{Tphi-icoupl}
\end{eqnarray}

\acknowledgments
Valuable discussions with David DiVincenzo are gratefully acknowledged.

\appendix

\section{Derivation of the Equations of Motion}
\label{deriv-eqmot}
This appendix contains the derivation of Eq.~(\ref{summary1}).
Note, first, that the externally applied magnetic flux ${\bf \Phi}_x$
only threads loops with a finite self-inductance (i.e., those pertaining to
a chord inductor, $K$), and not, e.g., the circuit loop formed by a junction $J$ and its
junction capacitance $C_J$, therefore 
${\bf \Phi}_x \equiv({\bf \Phi}_{CJ}^x,{\bf \Phi}_{C}^x,{\bf \Phi}_K^x)=({\bf 0},{\bf 0},{\bf \Phi}_K^x)$.
Using this fact and Eqs.~(\ref{IVct-2}) (capacitance part) and (\ref{Josephson-2}),
we obtain
\begin{eqnarray}
  \label{eqm-1}
  \frac{\Phi_0}{2\pi} {\bf F}_{JC}^T \dot\bphi  
          &=& {\bf V}_C - {\bf F}_{LC}^T {\bf V}_L - {\bf F}_{VC}^T {\bf V}_V - {\bf F}_{ZC}^T {\bf V}_Z\\
          &=& {\bf C}^{-1} {\bf Q}_C  - {\bf F}_{LC}^T \dot{\bf \Phi}_L 
              - {\bf F}_{VC}^T {\bf V}_V - {\bf F}_{ZC}^T {\bf Z}*{\bf I}_Z,
\nonumber
\end{eqnarray}
multiply this equation by ${\bf F}_{JC}{\bf C}$ and use Eq.~(\ref{IVct-1}) (impedance part),
with the result
\begin{eqnarray}
  \label{eqm-2}
  \frac{\Phi_0}{2\pi} {\bf F}_{JC}{\bf C}{\bf F}_{JC}^T \dot\bphi  
          &=& {\bf F}_{JC}{\bf Q}_C  
              - {\bf F}_{JC}{\bf C}{\bf F}_{LC}^T \dot{\bf \Phi}_L \nonumber\\
           & &  - {\bf F}_{JC}{\bf C}{\bf F}_{VC}^T {\bf V}_V  \nonumber\\
           & &  - {\bf F}_{JC}{\bf C}{\bf F}_{ZC}^T {\bf Z} {\bf F}_{ZC} * \dot{\bf Q}_C .
\end{eqnarray}
Then we make use of Eq.~(\ref{IVct-1}) (Josephson part) and obtain
\begin{eqnarray}
  \label{eqm-3}
  \frac{\Phi_0}{2\pi} {\bf C}_{\rm tot} \dot\bphi  
          &=& -{\bf Q}_J
              - {\bf F}_{JK}{\bf Q}_K
              - {\bf C}_{JL}\dot{\bf \Phi}_L \nonumber\\
          & & - {\bf C}_{JV}{\bf V}_V 
              - {\bf F}_{JC} {\bf C}_Z * {\bf V}_C,
\end{eqnarray}
where we have defined the frequency-dependent capacity
${\bf C}_Z(\omega) = i\omega {\bf C}{\bf F}_{ZC}^T {\bf Z}(\omega) {\bf F}_{ZC}{\bf C}$ and
\begin{eqnarray}
  {\bf C}_{\rm tot} &=& {\bf C}_J+{\bf F}_{JC}{\bf C}{\bf F}_{JC}^T,     \label{Ctot}\\
  {\bf C}_{JL}      &=& {\bf F}_{JC}{\bf C}{\bf F}_{LC}^T,     \label{CJL}\\
  {\bf C}_{JV}      &=& {\bf F}_{JC}{\bf C}{\bf F}_{VC}^T,     \label{CJV}
\end{eqnarray}
We find that ${\bf C}_Z(\omega)$ is a symmetric matrix since both ${\bf C}$ 
and ${\bf Z}$ are symmetric.
Using Eq.~(\ref{IVct-2}) (capacitance part) again, we obtain
\begin{equation}
  \label{eqm-4}
  {\bf F}_{LC}^T \dot{\bf \Phi}_L   =   {\bf C}^{-1} {\bf Q}_C  
                  -\frac{\Phi_0}{2\pi} {\bf F}_{JC}^T \dot\bphi
                  - {\bf F}_{VC}^T {\bf V}_V - {\bf F}_{ZC}^T {\bf Z}*{\bf I}_Z,
\end{equation}
which we multiply with ${\bf F}_{LC}{\bf C}$, with the result
\begin{eqnarray}
  \label{eqm-5}
  {\bf F}_{LC}{\bf C}{\bf F}_{LC}^T \dot{\bf \Phi}_L   
             &=&   {\bf F}_{LC}{\bf Q}_C  
                  - \frac{\Phi_0}{2\pi} {\bf F}_{LC}{\bf C}{\bf F}_{JC}^T \dot\bphi\nonumber\\
              & & - {\bf F}_{LC}{\bf C}{\bf F}_{VC}^T {\bf V}_V \nonumber\\
              & & - {\bf F}_{LC}{\bf C}{\bf F}_{ZC}^T {\bf Z}{\bf F}_{ZC} * \dot{\bf Q}_C.
\end{eqnarray}
With the definitions ${\bf C}_L = {\bf F}_{LC}{\bf C}{\bf F}_{LC}^T$
and ${\bf C}_{LV} = {\bf F}_{LC}{\bf C}{\bf F}_{VC}^T$, we find
\begin{eqnarray}
  \label{eqm-6}
  {\bf C}_L \dot{\bf \Phi}_L   
             &=&    -{\bf Q}_L - {\bf F}_{LK}{\bf Q}_K
                  - \frac{\Phi_0}{2\pi} {\bf C}_{JL}^T \dot\bphi\nonumber\\
              & & - {\bf C}_{LV} {\bf V}_V
                  - {\bf F}_{LC}{\bf C}_Z * {\bf V}_C.
\end{eqnarray}
Equations (\ref{eqm-3}) and (\ref{eqm-6}) are rewritten
in a more compact form in Eq.~(\ref{summary1}).

\section{System-bath dynamics}
\label{CLderivation}
In this section, the form of the system-bath coupling operator ${\cal H}_{SB}$ and its spectral
density $J(\omega)$, Eqs.~(\ref{eq:HSB}) and (\ref{eq:JK}), are derived in detail.

We first inspect the Hamilton equations for the bath coordinates,
\begin{eqnarray}
  \dot x_\alpha  &=&  \frac{\partial {\cal H}}{\partial p_\alpha} = \frac{p_\alpha}{m_\alpha},\label{xt}\\
  \dot p_\alpha  &=&  -\frac{\partial {\cal H}}{\partial x_\alpha} 
                  = -m_\alpha \omega_\alpha^2 x_\alpha - c_\alpha \bar{\bf m}\cdot {\cal C}^{-1}{\bf Q},
\label{pt}
\end{eqnarray}
then take their derivative with respect to time, and solve them in Fourier space.
We obtain
\begin{eqnarray}
  x_\alpha(\omega) &=& \frac{c_\alpha \bar{\bf m}\cdot{\cal C}^{-1}{\bf Q}}{m_\alpha \left( \omega^2 -\omega_\alpha^2\right)},\label{xomega}\\
  p_\alpha(\omega) &=& m_\alpha i\omega x_\alpha(\omega) 
                    = \frac{i\omega c_\alpha \bar{\bf m}\cdot{\cal C}^{-1}{\bf Q}}{\omega^2 -\omega_\alpha^2}.\label{pomega}
\end{eqnarray}
Next, we look at the Hamilton equations for the system coordinates,
\begin{eqnarray}
  \dot {\bf\Phi}  &=& \frac{\partial {\cal H}}{\partial {\bf Q}} 
                   = {\cal C}^{-1}\left({\bf Q} 
                   + \bar{\bf m}\sum_\alpha c_\alpha x_\alpha\right),  \label{Phit}\\
  \dot {\bf Q}    &=&  -\frac{\partial {\cal H}}{\partial {\bf \Phi}} 
                   =   -\frac{\partial U}{\partial {\bf \Phi}}.\label{Qt}
\end{eqnarray}
Combining Eqs.~(\ref{Phit}) and (\ref{Qt}) with Eqs.~(\ref{xt}) and (\ref{pomega}), we obtain
\begin{equation}
\label{Phitt}
  {\cal C} \ddot {\bf \Phi}  =  -\frac{\partial U}{\partial {\bf \Phi}} 
                             +\bar{\bf m}\sum_\alpha c_\alpha \frac{p_\alpha}{m_\alpha}
   =  -\frac{\partial U}{\partial\bphi} - K*\bar{\bf m}(\bar{\bf m}\cdot {\cal C}^{-1}\dot{\bf Q})
\end{equation}
where
\begin{equation}
  K(\omega) = -\sum_\alpha \frac{c_\alpha^2}{\omega^2-\omega_\alpha^2}
\end{equation}
directly determines the bath spectral density
\begin{equation}
  J(\omega) =  \frac{\pi}{2}\sum_\alpha \frac{c_\alpha^2}{m_\alpha \omega_\alpha}\delta(\omega-\omega_\alpha) = -{\rm Im} K(\omega).
\end{equation}
By comparing Eq.~(\ref{Phitt}) with Eq.~(\ref{eq:eq-mot}), we find
\begin{equation}
  {\cal C}_d(\omega)  = K(\omega)\bar{\bf m}\bar{\bf m}^T.
\end{equation}


\begin{thebibliography}{99}
\bibitem{fn-phase}
A third variant of SC qubits, operating in the regime $E_J\gg E_C$, are the
phase qubits \cite{Ioffe,Martinis02}.  For the Josephson phase qubit \cite{Martinis02},
decoherence due to bias noise and junction resonators 
was studied in \cite{Martinis03,Simmonds04}.
\bibitem{fn1}
Current bias sources shunted with an impedance $Z$ can be included as well, 
using Norton's theorem to replace them with a voltage source with the
same impedance $Z$ in series, with applied voltage $V=Z(0)I$ for a dc 
current bias $I$.
\end{thebibliography}
\end{document}